\documentclass[12pt]{article}

\usepackage{graphicx}

\newcommand{\rf}[1]{(\ref{#1})}
\newcommand{\beq}{\begin{equation}}

\newcommand{\eeq}{\end{equation}}
\newcommand{\bea}{\begin{eqnarray}}
\newcommand{\eea}{\end{eqnarray}}

\newcommand{\e}{\mbox{e}}
\renewcommand{\d}{\mbox{d}}
\newcommand{\g}{\gamma}

\newcommand{\lam}{\lambda}
\newcommand{\La}{\Lambda}
\renewcommand{\b}{\beta}
\renewcommand{\a}{\alpha}
\newcommand{\n}{\nu}

\newcommand{\ep}{\varepsilon}

\newcommand{\om}{\omega}

\newcommand{\del}{\delta}
\newcommand{\Del}{\Delta}

\newcommand{\kp}{\kappa}

\newcommand{\ra}{\rangle}
\newcommand{\la}{\langle}

\newcommand{\mi}{\!-\!}

\newcommand{\cD}{{\cal D}}

\newcommand{\tV}{{\tilde{V}}}

\newcommand{\tc}{{\tilde{c}}}

\newcommand{\tN}{{\tilde{N}}}

\newcommand{\hG}{{\hat{G}}}

\begin{document}

\begin{center}
\vspace{24pt}
{ \large \bf CDT and the Search for a Theory of Quantum Gravity}

\vspace{30pt}

{\sl J. Ambj\o rn}$\,^{a,c}$,
{\sl A. G\"{o}rlich}$\,^{a,b}$
{\sl J. Jurkiewicz}$\,^{b}$,
and {\sl R. Loll}$\,^{c,d}$

\vspace{24pt}
{\footnotesize

$^a$~The Niels Bohr Institute, Copenhagen University\\
Blegdamsvej 17, DK-2100 Copenhagen \O , Denmark.\\
{\it email: ambjorn@nbi.dk, goerlich@nbi.dk}\\

\vspace{10pt}

$^b$Institute of Physics, Jagellonian University,\\
Reymonta 4,  PL 30-059 Krakow, Poland.\\
{\it email: jerzy.jurkiewi@uj.edu.pl} 
\vspace{10pt}

$^c$Radboud University Nijmegen,\\ 
Institute for Mathematics, Astrophysics and Particle Physics,\\ 
Heyendaalseweg 135, 6525 AJ Nijmegen, The Netherlands. \\
{\it email: r.loll@science.ru.nl}

\vspace{10pt}

$^d$Perimeter Institute for Theoretical Physics,\\
31 Caroline St N, Waterloo, Ontario N2L 2Y5, Canada.

}

\end{center}

\vspace{30pt}

\begin{center}
{\bf Abstract}
\end{center}

Causal Dynamical Triangulations provide a non-perturbative 
regularization of a theory of quantum gravity. We describe how 
this approach connects with the asymptotic safety program and
Ho\v rava-Lifshitz gravity theory, and present the most recent 
results from computer simulations.

\vspace{30pt}

{\sl keywords:} MG13 Proceedings. Plenary talk. Quantum gravity.

\newpage

\section{Introduction}\label{introduction}

We do not know if there exists a theory of quantum 
gravity and precisely how we should think about 
quantizing geometry. The amazing successes of quantum 
mechanics and quantum field theory have been obtained
in the rigid background of flat spacetime. Can we 
extend the realm of quantization to include the geometry
of spacetime itself in a consistent way? Even if the answer is
affirmative (as will be argued 
shortly), we are still faced with the problem that 
starting out with general relativity (GR) and expanding around a 
fixed classical background geometry, the  
gravitational fluctuations are non-renormalizable in a 
conventional field-theoretical sense. 

How should we deal with the non-renormalizable aspects? The logic behind string theory
is in a sense in line with how  
non-renormalizable theories have been handled
until now. Two examples in point are the four-fermion model
of the weak interactions and the non-linear sigma model
of pion interactions. In both cases new degrees of freedom were introduced,
which made the more fundamental, underlying field theories 
renormalizable. The apparent non-renormalizability of the models was
due to our incomplete understanding of the underlying short-distance 
physics. String theory introduces an infinite number of 
new field degrees of freedom in order to tame the UV divergences 
of the gravitational field. This solution to the 
UV problem has a certain elegance, since it can be described as moving from 
zero-dimensional point particles to one-dimensional strings in an
(almost) unique way, dictated by relativistic invariance. When 
the string is expanded into modes, one is led 
to an infinite set of fields. However, despite the presence of 
an enormous number of degrees 
of freedom string theory has
had difficulties reproducing anything like the universe we observe. 
Loop quantum gravity and its 
ramifications constitute other attempts to define a theory of quantum 
gravity \cite{cambridge}. In loop quantum gravity one insists that holonomies 
are quantum objects which are finite. This leads to a somewhat 
non-standard quantization with non-separable Hilbert spaces,
and it is in general difficult to show that one obtains 
a $\hbar \to 0$ limit which can be identified with classical GR.

The asymptotic safety program \cite{weinberg} tries to address the problem of 
non-renormalizability in a more mundane way, using only 
ordinary quantum field theory. It appeals to the Wilsonian
concept of renormalizability, where the UV or IR behaviour 
of a quantum field theory are linked to the existence of fixed 
points of the renormalization group. Accordingly, in this framework 
one conjectures that the behaviour of quantum gravity at high energies is governed by a 
non-Gaussian fixed point in the UV, which has
properties similar to the Gaussian UV fixed point known from renormalizable
quantum field theories. In a renormalizable theory with a
Gaussian UV fixed point, sufficiently 
close to it there are only a finite 
number of independent operators that take us away from the fixed point
when we integrate out high-frequency modes. Furthermore, the coupling 
constants associated with these operators scale to zero at the fixed point. 
For a Gaussian UV fixed point in flat spacetime we can 
(except for possible complications with gauge invariance etc.) obtain 
the possible operators by power counting. The associated coupling constants 
are in principle the only ones that need fine-tuning in order to reach the 
fixed point. One can use perturbation theory to check this 
picture when one is close to the Gaussian fixed point.
Asymptotic safety assumes that a similar scenario is valid for a putative 
non-Gaussian fixed point, the only problem being that we currently
have no examples of such non-Gaussian UV fixed points in four-dimensional
conventional quantum field theory. 
Since the fixed point in question is non-Gaussian, 
we do not have perturbation theory available to investigate it.
It is nevertheless a logical possibility that 
such a fixed point exists precisely in the case of quantum gravity.
Over the last 20 years investigations using improved renormalization
group techniques\cite{exactRG}, as well as 2+$\ep$ dimension 
expansions\cite{kawai} have produced some evidence for the existence
of such a UV fixed point  for quantum gravity. 

Finally, a simple 
way to address the non-renormalizability issue using only conventional 
quantum field theory has more recently been suggested by Ho\v rava and 
since been dubbed Ho\v rava-Lifshitz gravity\cite{horava}. 
It emphasizes unitarity and 
the requirement that the theory should be renormalizable. It achieves
this by insisting on at most second-order time derivatives in the action
but allowing for higher-order spatial derivatives. The price one pays 
is that the theory violates Lorentz invariance at short distances
where the higher-derivative terms are important, the hope being that --
in agreement with observations --
the Lorentz invariance is approximately restored at large distances. 

Causal dynamical triangulations (CDT) is a lattice regularization
of quantum gravity\cite{al,ajl1}. 
It provides us with a non-perturbative definition
of the path integral of quantum gravity, where the length of the 
lattice links $a$ acts as a UV cut-off. Formally, the 
continuum limit is obtained when $a \to 0$. As a lattice theory 
it fits naturally into a Wilsonian framework and can therefore 
be thought of as an independent way of investigating the asymptotic safety scenario;
one has a phase diagram in terms of the bare coupling constants and 
looks for second-order phase transition points which can serve as the 
non-Gaussian UV fixed points of asymptotic safety. Having identified 
such fixed points one can study the renormalization group flow close
to them. At the same time, CDT is well suited to serve 
as a regularization of Ho\v rava-Lifshitz (HL) gravity models,
since both in CDT and HL gravity one integrates in the path integral over 
geometries which have a preferred time foliation. As will be discussed below, the CDT  
phase diagram also has many similarities with a Lifshitz phase diagram.   
  
Before entering the detailed discussion of CDT results, let us discuss
three key issues:
\begin{itemize}
\item[(1)] Does it make sense to talk about a quantum theory of gravity, which includes 
fluctuating geometries, when conventional quantization is always 
performed with reference to a fixed geometry? 
\item[(2)] Does it make 
sense to consider a lattice regularization of a diffeomorphism-invariant theory? 
\item[(3)] Does it make sense to 
talk about a Wilsonian framework in a diffeomorphism-invariant theory,
where one has not even defined what is meant by a (diffeomorphism-invariant)
correlation length? 
\end{itemize}
As it happens, all of the above questions can be answered affirmatively without
addressing the more difficult question of whether there exists a quantum gravity
theory in four-dimensional spacetime. In two dimensions, gravity becomes 
a very simple theory, since the integral of the scalar curvature is 
topological. Thus there {\it is} no dynamical action. Nevertheless
the theory has a non-trivial partition function, and produces non-trivial amplitudes 
for universes with one boundary geometry ``propagating'' into another 
boundary geometry and for matter fields living on such fluctuating
geometries. Of course, this theory has no propagating local gravitational 
degrees of freedom. However, the conceptual questions mentioned above 
do not really refer to the propagation of gravitational degrees 
of freedom, but to the fact that we are dealing with fluctuating 
geometry, in a setting where we only want to ask diffeomorphism-invariant questions. 
Viewed in this perspective, one could even say that the two-dimensional theory provides us 
with the ultimate test, since there hardly {\it is} any classical action (only
the cosmological term). If we think about the quantum theory using 
the path integral as a sum over spacetime histories, we are dealing with a
situation that is ``maximally quantum", in the sense that each configuration 
in the path integral has the same 
weight. This corresponds formally to the limit 
$\hbar \to \infty$, i.e.\ the ultimate quantum limit.  

So-called 2d Liouville quantum gravity can be solved analytically,
both using canonical quantization and the path integral. It {\it is} 
the quantum theory of fluctuating two-dimensional geometries with fixed topology,
and therefore answers (1) above. The same theory can be regularized using lattices,
using the formalism called ``dynamical triangulations'' (DT). In it 
one performs the path integral in the Euclidean sector
by using equilateral triangles as building blocks, and gluing them 
together in all possible ways compatible with the given topology. In this 
way the path integral becomes the sum over a class of piecewise linear 
geometries, whose geometry is determined entirely by the way the building 
blocks are glued together, justifying the name ``dynamical triangulations''. 
The link length $a$ of the building blocks acts as a UV cut-off. 
Somewhat surprisingly, one can solve this lattice theory analytically
and in the limit $a\to 0$ recover the continuum quantum Liouville 
results. In other words, there exists a lattice regularization where one
sums over spacetime {\it geometries} (as opposed to spacetime {\it metrics}),  
no gauge-fixing is needed (and therefore no issue of diffeomorphism-invariance 
arises), and the $a \to 0$ limit leads to 
a diffeomorphism-invariant theory, namely, quantum Liouville theory.   

The lattice theory also fits beautifully into a Wilsonian framework,
thereby answering question (3) above. To start with, it has {\it universality} in 
the sense that we are not restricted to using equilateral triangles 
as building blocks. Using instead almost any ensemble of polygons with 
side-length $a$, and gluing them together with almost any positive weights,
the continuum limit $a\to 0$ will still be the same. By adding
suitable higher-curvature action terms to this theory of planar graphs
one again obtains the same continuum limit, demonstrating
a Wilsonian universality with respect to both regularization and the choice 
of action. 

Finally, is it possible 
to define the concept of diffeomorphism-invariant correlators depending
on a diffeomorphism-invariant correlation length? In two dimensions, where we 
have no propagating gravitational field degrees of freedom, one has 
to add matter fields to the fluctuating geometries to answer this
question. This can be done, and the answer is again in the affirmative
\cite{correlators}. To be more explicit, let us consider a matter field
$\phi(x)$, whose dynamics is governed by some diffeomorphism-invariant 
action. Clearly it makes no sense to 
consider the correlation between two local fields $\phi(x)$ and $\phi(y)$, since $x$ and 
$y$ are just coordinates. In addition, since the geometry of spacetime 
is fluctuating, the geodesic distance between $x$ and $y$ will depend
on the geometry, and we are summing over these geometries. How 
{\it do} we then define a meaningful diffeomorphism-invariant correlation between fields?
In flat $d$-dimensional spacetime, let us rewrite the correlator of  
the scalar field $\phi(x)$ in the form
\beq\label{h7}
\la \phi \phi (R) \ra_V \equiv 
\frac{1}{V} \;\frac{1}{s(R)} \int \cD \phi\, e^{-S[\phi]} 
\int\!\! d^dx \!\int \!\! d^dy
\;  \phi(x) \phi(y)   
\;\del(R \mi |x-y|).   
\eeq
As indicated, this expression depends on a chosen distance $R$, 
but no longer on specific 
points $x$ and $y$, which instead are integrated over. 
The integrand can be read 
``from right to left" as first averaging 
over all points $y$ at a distance $R$ from some fixed point $x$, normalized
by the volume $s(R)$ of the spherical shell of radius 
$R$, and then averaging 
over all points $x$, normalized
by the total volume $V$ of spacetime. 
We assume translational and 
rotational invariance of the theory and that $V$ is so large that 
we can ignore any boundary effects related to a finite volume.
This definition of a correlator is of course non-local, but unlike
the underlying locally defined correlator has a straightforward 
diffeomorphism-invariant 
generalization to the case where gravity is dynamical, namely,
\bea
\la \phi \phi (R) \ra_V &\equiv& 
\frac{1}{V}\int \cD [g]\, \int \cD_{[g]} \phi \;\e^{-S[g,\phi]} 
\;\del\Big(V-\!\!\int \!\! d^dx \sqrt{\det g} \Big)\times \nonumber \\  
&&  
\int\!\! d^dx \!\int \!\! d^dy\, \;\frac{\sqrt{\det g(x)} \, 
\sqrt{\det g(y)}}{s_{[g]}(y,R)}
\; \phi(x) \phi(y)  
\;\del(R \mi D_{[g]}(x,y)),  
\label{h8}
\eea
which now includes a functional integration over 
geometries. The geometry corresponding to a metric $g_{ij}$ 
is denoted by $[g]$ and the
dependence of the action, measures, distances and volumes refers
to the specific geometry $[g]$, which is finally integrated over,
but with $R$ and $V$ kept fixed.
It can be shown that this definition allows us to think about 
correlators in the standard way. Most importantly in this 
context is the fact that the Wilsonian concept of a divergent correlation
length when approaching a second-order phase transition --
key to the universality of the continuum limit -- is still true
when integrating over fluctuating geometries\cite{correlators}.

\section{Causal Dynamical Triangulations}

\begin{figure}[t]
\centerline{\includegraphics[width=4in]{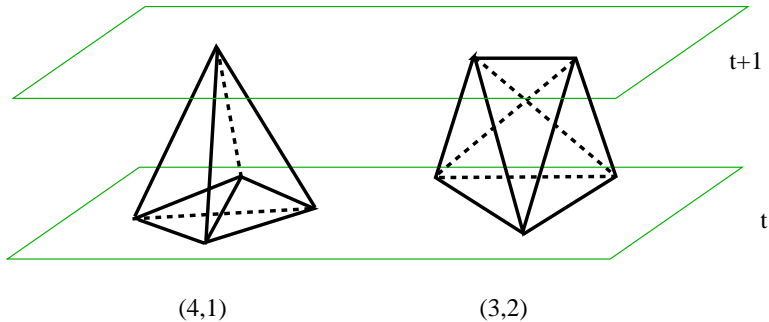}}
\caption{A triangulation in CDT consists of four-dimensional triangulated
layers assembled from (4,1)- and (3,2)-simplices, interpolating between
adjacent integer constant-time slices (left), which in turn are triangulations 
of $S^3$ in terms of equilateral
tetrahedra. Each purely spatial tetrahedron at 
time $t$ forms the interface between two
(4,1)-simplices, one in the interval $[t-1,1]$, 
and the other in $[t,t+1]$, as
illustrated on the right. Although a 
(3,2)-simplex shares none of the five tetrahedra on its
surface with a constant-time slice 
(the tetrahedra are all Lorentzian), it is nevertheless
needed in addition to the (4,1)-building block to obtain 
simplicial manifolds with a well-defined causal structure.}
\label{fig1}
\end{figure}
  
The CDT lattice regularization differs in a crucial way from the 
DT lattice regularization discussed in the Introduction.
While it uses the same action (the discretized Einstein-Hilbert action,
in the form of the so-called Regge action\cite{regge})
as DT, it uses only triangulated manifolds with the topology of a direct
product $[0,1]\times {}^3\Sigma$, and subject to a preferred  
proper-time foliation in the time direction.
The motivation for this is twofold. While DT is 
inherently Euclidean in its set-up, CDT is an attempt to define the 
regularized path integral in spacetimes with Lorentzian signature.
Following earlier ideas\cite{teitelboim} 
we insist on including in the path integral 
only  locally causal geometries, whose light cone structure 
is everywhere non-degenerate. 
CDT is an attempt 
to implement such a path integral, defined as  a
lattice regularization, much in the same way as was done using DT.
(We refer the reader to a comprehensive recent 
review \cite{physrep} for details.) Like the DT geometries, the  CDT geometries 
are piecewise linear geometries constructed by gluing together a few 
standard building blocks, but such that they respect the proper-time 
foliation. In this construction we have space-like lattice links, 
to which we assign a length $ds^2 = a_s^2= a^2$, 
and time-like lattice links, to
which we assign a proper-time length $ds^2= a^2_t = -\a a^2$, where 
$a$ and $a_s$ are positive. This is illustrated in Fig.\ \ref{fig1}
in the case of four-dimensional spacetime, where four standard building 
blocks are needed, the so-called (4,1)- and (3,2)-building blocks 
shown in the figure and time-reversed (1,4)- and (2,3)-building 
blocks, the numbers referring to the number of vertices of the four-dimensional
simplex at spatial slice $t$ and spatial slice $t+1$. The spatial slice at any 
$t$ is assumed to have the topology ${}^3\Sigma=S^3$ (we assume that space is 
compact, and have chosen the simplest topology), and is triangulated
in terms of equilateral tetrahedra of link length $a_s$. At discrete times $t$ and $t+1$ 
we thus have two purely spatial triangulations of $S^3$. Fig.\ \ref{fig1} illustrates
how to fill in the four-dimensional ``slab'' between the two $S^3$-triangulations 
with four-simplices, 
preserving the topology $[0,1]\times S^3$. In the path integral 
we will sum over all triangulations of the three-sphere at $t$, all triangulations
of the three-sphere at $t+1$, etc., 
and over all four-dimensional triangulations of the slabs
compatible with the given choices of the spherical triangulations at $t$, 
$t+1$, $\dots$ . For each spacetime geometry thus obtained we 
can perform a rotation to Euclidean signature by rotating 
$\a$ from positive to negative values in the lower-half complex plane. 
The corresponding Regge action also rotates in the way expected
for a rotation from Lorentzian to Euclidean signature, that is,
\beq\label{j1}
i S_L(\a) = -S_E(-\a).
\eeq  
The constraint of local causality is not a topological constraint, and 
after rotation to Euclidean signature leads to a summation
over a restricted class of Euclidean geometries. The corresponding theory
is therefore potentially different from 
``Euclidean quantum gravity'' \`a la Hawking. This brings 
us to the second motivation for replacing the DT 
with the CDT regularization in spacetime dimensions $d$ larger 
than two. If one restricts oneself to the 
Euclidean Einstein-Hilbert action in Regge form, the DT lattice theory
has no continuum limit. Studying the theory with Monte Carlo simulations, 
one finds a phase transition as a function of the bare 
gravitational coupling constant, but it is of first order. 
Such a first-order transition cannot be used 
to define a continuum quantum field 
theory. As discussed in the Introduction, we need a second-order transition 
to which we can associate a divergent correlation length. 
The situation will be different when we use the CDT regularization.

\section{CDT phase structure}

We will discuss here the four-dimensional CDT theory. We rotate it 
to Euclidean signature as described above. Because the four-dimensional theory 
cannot be solved 
analytically, we need to rotate it to Euclidean signature to be able to study it
by Monte Carlo simulations. 

Let us first write down the regularized Euclidean Einstein-Hilbert action
for a CDT configuration. As mentioned earlier, we use the so-called
Regge action, which can be used for 
any piecewise linear geometry. For a $d$-dimensional 
triangulation the curvature will be located at  the $(d-2)$-dimensional 
subsimplices, which implies that in four dimensions the curvature
is concentrated at the two-dimensional subsimplices (the triangles). 
In the case of CDT the Regge action becomes exceedingly  
simple, because we are using fixed building blocks to construct the 
geometries. The curvature depends only on how we glue
these building blocks together. After using various identities relating 
the number of subsimplices and the order of vertices, links and 
triangles, the action will only depend on the total number $N_0$ of vertices,
the total numbers $N_4^{(4,1)}$ and $N_4^{(3,2)}$ of (4,1)- and (3,2)-simplices 
and the parameter $\a$,
\bea\label{h19}
S_E[-\a;T] &=& -(\kp_0+6 \Del)N_0(T) +
\kp_4\Big(N_4^{(3,2)}(T)+N_4^{(4,1)}(T)\Big)  \\ 
&& + \Del \Big(N_4^{(3,2)}(T)+2N_4^{(4,1)}(T)\Big),
\nonumber
\eea 
where the asymmetry parameter $\Del$ \index{Asymmetry Parameter $\Del$}
is a function of $\a$ such that $\Del(\a\!=\! 1) =0$.
In this formula $\kp_0 \propto 1/G_0$, the bare inverse gravitational coupling 
constant, while $\kp_4$ can be more or less identified with $\La_0/G_0$,
$\La_0$ being the bare cosmological coupling constant. The CDT partition
function is given by 
\beq\label{j4}
Z(\kp_0,\kp_4,\Del)= \sum_T \e^{-S_E[T]},
\eeq
where one sums over four-dimensional CDT triangulations of the 
kind described in the previous Section. We should point out that
formula \rf{j4} contains a small subtlety. In the first place, one would 
not consider $\Del$ as a coupling constant. It was merely 
chosen to parametrize the asymmetry between the links in spatial and temporal 
directions, but the classical Einstein-Hilbert action was adjusted 
precisely to take this into account. However, it turns out 
that in the region of bare coupling-constant space where 
we observe a potentially interesting phase structure, the {\it entropy}
of  configurations with the same action is as important as 
the action itself\footnote{By entropy we here simply mean the 
relation $S = \ln ({\rm number~of~configurations})$.}. 
Although the entropy is independent of 
$\Del$, the real quantum effective action becomes in this way
a function of $\Del$. In the classical limit the contribution of the (bare) action 
will always be much more important than the entropy term, simply because 
the action is multiplied by $1/\hbar$, while this is not the case 
for the entropy term. However, the interesting Planckian physics we observe is 
of course {\it not} in this semiclassical limit. It follows that we
have a theory depending on
{\it three} coupling constants, $\kp_0$, $\kp_4$ and $\Del$. 
For technical convenience,  we keep the four-volume fixed during
the Monte Carlo simulations, which effectively removes $\kp_4$ as a 
coupling constant. We are thus left with two coupling constants, $\kp_0$ and $\Del$. 

\begin{figure}[t]
\centerline{\includegraphics[width=4in]{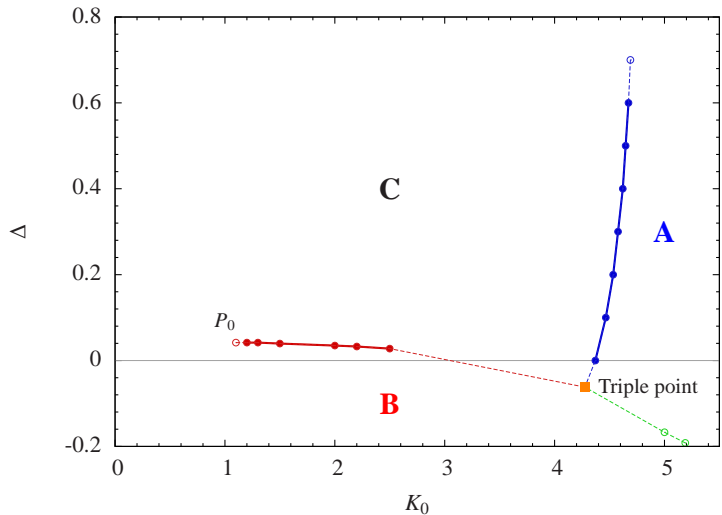}}
\caption{The phase diagram of four-dimensional CDT.}
\label{fig2}
\end{figure}

The numerical set-up is as follows: we choose a four-volume 
(the discrete number of four-simplices) and a large proper-time extent $t$ (the discrete number 
of lattice time steps), and perform the Monte Carlo simulations 
for a given choice of $\kp_0$ and $\Del$. We can measure the 
three-volume distribution as a function of proper time $t'$ between 0 and $t$, and deduce
from the measurements that there are
three qualitatively different types of  three-volume profiles $N_3(t')$,
depending on the choice of $\kp_0$ and $\Del$. Here $N_3(t')$ denotes  
the number of tetrahedra forming a triangulation  of the 
spatial slice at time $t'$. The different profiles correspond 
to different phases, which we have denoted by A, B and C \cite{ABC}. 
The corresponding phase diagram 
is shown in Fig.\ \ref{fig2} (we refer to elsewhere\cite{phasediagram} for details). 
In phase B we observe a {\it collapse} of the four-dimensional universe 
to a three-dimension one, in the sense that all of the three-volume 
$N_3(t')$, $t'\in [0,t]$, is located at a {\it single} time.
The situation in phase A is completely opposite. The three-volume $N_3(t')$ is 
spread out over the whole time interval $[0,t]$, and the magnitude of $N_3(t')$
is close to being a random variable, with little correlation with the neighbouring 
three-volumes $N_3(t'+1)$ and $N_3(t'-1)$ (at least, the correlation is rather short-ranged).
Finally, in phase C the situation is again very different. There is no 
collapse of three-volume around a single spatial slice. Instead, the three-volumes $N_3(t')$ 
at adjacent times are highly correlated, resulting in a genuinely
four-dimensional universe. The universe 
has a definite extension in the time direction, independent of the choice of $t$, as long as 
$t$ is sufficiently large compared to the given choice of $N_4$. Along the 
remainder of the $t$-axis one finds only vanishing three-volume (to be precise, one finds 
three-volumes which are close to the minimal cut-off size of 5 tetrahedra, since we
do not allow the computer algorithm to shrink the three-volume to strictly
zero). The actual time extension of the universe depends on the 
chosen $N_4$ and scales proportional to $N_4^{1/4}$ as one would na\"ively expect.
Similarly, the non-zero three-volumes $N_3(t')$ scale proportional to  $N_4^{3/4}$. 
This is the reason why we claim that the universe is four-dimensional\cite{ABC}. 
At first sight it may seem a triviality that 
the universe should scale in this fashion,
but it is in fact highly non-trivial. Although the elementary
building blocks are four-dimensional, nothing tells us that 
they will line up to form a four-dimensional object with 
approximately $N_4^{1/4}$ building blocks in each 
direction. Since absolutely no ``background'' geometry has been put 
in by hand to ensure the four-dimensionality of the resulting ``quantum spacetime",
this will in general not happen at any scale, as exemplified by the situation in phases A and B. 
Furthermore, as we have already noted and 
as will be discussed later, in phase C we are far from any choice 
of the bare coupling constants where the classical action dominates.

We thus have identified three phases with very different geometric large-scale characteristics.
Associated with these phases are phase transition lines, as shown in Fig.\ \ref{fig2}.
We have investigated the order of these transitions, with the result 
that the transition A-C is first order, while the transition B-C
is second order\cite{jordan}.  
(We are currently not particularly interested in the transition 
A-B, since geometry in neither phase A nor B seems relevant as 
a model for a four-dimensional universe.)
\begin{figure}[t]
\centerline{\includegraphics[width=4in]{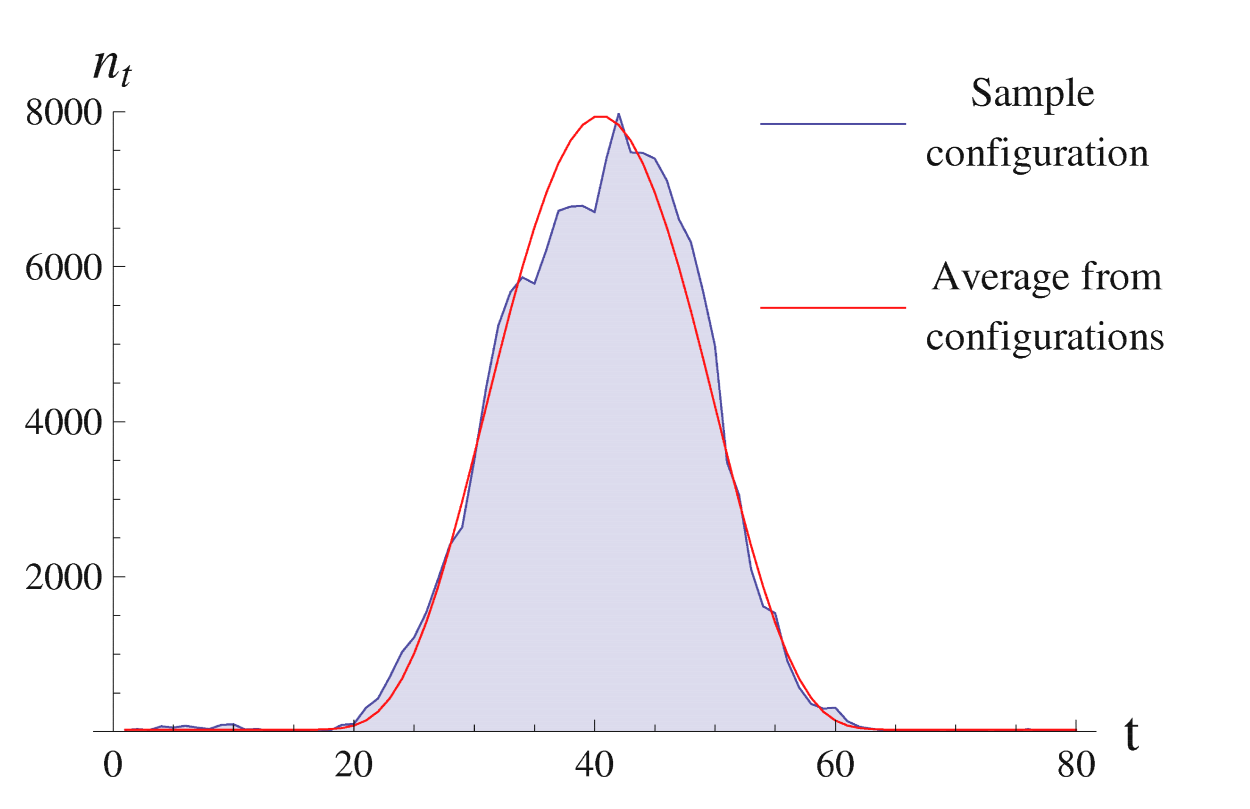}}
\caption{A typical three-volume profile (and the average profile)
for some choice of coupling constants in phase C.}
\label{fig3}
\end{figure}

Following the Wilsonian logic on how to take the continuum limit,
our main interest will be focused on the B-C phase transition line. Can it 
be used to define a UV limit of quantum gravity? Before addressing this
point, we will take a closer look at the situation in phase C, where 
we seemingly observe four-dimensional universes.

\section{The physics of phase C}

\begin{figure}[t]
\centerline{\includegraphics[width=4in]{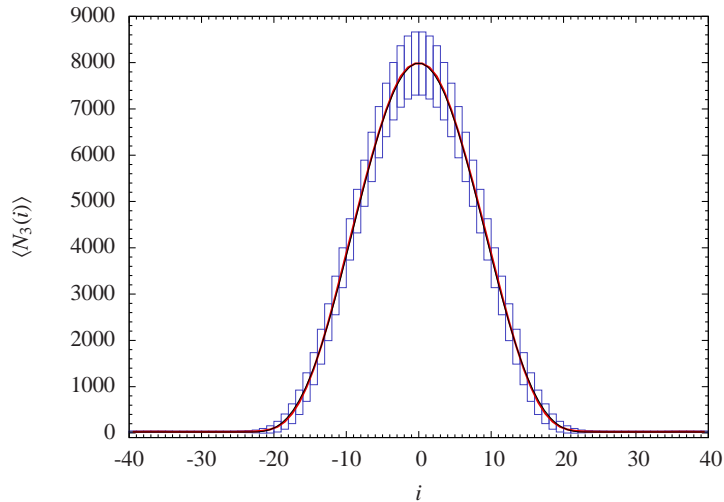}}
\caption{The average three-volume profile and the fluctuations
for $N_4= 360.000$, at $\kp_0=2.2$ and $\Del=0.6$.}
\label{fig4}
\end{figure}

A typical three-volume profile in phase C is shown in Fig.\ \ref{fig3}.
Fig.\ \ref{fig4} highlights the average value of the 
profile and the size of typical fluctuations. The data are 
collected for fixed values of the coupling constants and 
fixed $N_4$. For the average profile there 
is a perfect scaling $\propto N_4^{3/4}$ of the height and $\propto N_4^{1/4}$ 
of the width of the profile. However, the fluctuations
of the height of the profile  
scale like $N_4^{1/2}$. This means that for fixed values of the bare 
coupling constants, taking $N_4\to \infty$, the relative fluctuations
will go to zero and we will have a well-defined limit.
For fixed values of the bare coupling 
constants, the average profile can be fitted perfectly to the formula
\beq\label{h24}
 \la N_3(i)\ra  \propto
N_4^{{3}/{4}}
\cos^3 \left(\frac{i}{s_0 N_4^{1/4}}\right),
\eeq
where $i$ denotes (integer) lattice time, $N_4$ the total 
number of four-simplices\footnote{For fixed values 
of the coupling constants, we can use for $N_4$ either the total number 
of four-simplices or the total number $(4,1)$-simplices, say. The numbers
will be proportional, the constant of proportionality changing 
somewhat with the coupling constants $\kp_0$ and $\Del$.} 
and $N_3(i)$ the number of tetrahedra 
at time $i$ \cite{s4a}, and $s_0$ is a constant 
which will depend on the choice of bare coupling constants 
(the formula is of course not valid in the ``stalk'', where
$N_3(i) \approx 5$.). 

Can the functional form of the expectation value found in \rf{h24} be obtained 
directly from an action principle? The answer is yes \cite{semi}.
The minisuperspace approach 
to quantum gravity by Hartle and Hawking\cite{hawking}, which only involves 
the scale factor of the universe (equivalently, the three-volume)
reduces the Einstein-Hilbert action to the 
``effective'' action 
\beq\label{h25}
S_{\mathit ef\! f}= -\frac{1}{24\pi G} \int d t 
\left( \frac{ \dot{V_3}^2(t)}{V_3(t)}+k_2 V_3^{1/3}(t)
-\lam V_3(t)\right),
\eeq
where $t$ denotes proper time, $k_2$ is a numerical constant
and $\lam$ is a Lagrange multiplier ({\it not} a cosmological constant), 
because the total four-volume $V_4$ is kept fixed in the simulations.
The classical solution of the equations of motion 
corresponding to $S_{ef\! f}$ is precisely 
the $\cos^3 (t/V_4^{1/4})$ referred to in eq.\ \rf{h24}.  

From the computer simulations and the measurements of 
$\la N_3(i) \ra$ and $\la N_3(i)N_3(j) \ra$ it is possible
to reconstruct the effective discretized action leading 
to \rf{h24}. It has the form\cite{s4a} 
\beq\label{h26}
S_{discr} =
k_1 \sum_i \left(\frac{(N_3(i+1)-N_3(i))^2}{N_3(i)}+
\tilde{k}_2 N_3^{1/3}(i)-\tilde{\lam} N_3(i)\right).
\eeq
This is precisely the discretized version of \rf{h25}, {\it 
except for an overall sign}. This overall sign does not affect the 
equations of motion derived from the actions, but it shows that in phase C
we are far away from any semiclassical limit, even if the 
effective outcome (the average universe) is similar. Since the effective
action \rf{h26} comes from combining the discretized, bare Einstein-Hilbert action
and the entropy of configurations, as discussed earlier, it is clear
that entropy plays an important role in phase C. 

Finally, from the effective action \rf{h26}, and comparing to \rf{h25},
it is natural to consider $k_1$, which we can actually measure, 
as the dimensionless coupling constant
\beq\label{j10}
k_1 \propto  a^2/G(a),
\eeq   
where we have assumed a scale-dependence of the gravitational 
coupling constant $G$, using as scale parameter the lattice 
cut-off $a$. 
In the asymptotic safety scenario, $k_1$ should run to a constant
different from zero when we approach a non-trivial UV fixed point,
while it should go to zero in the IR, where $G(a) \to G$, the ordinary 
gravitational coupling constant. We have measured $k_1(\kp_0,\Del)$, and there
is a clear indication that it goes to zero when we keep $\Del$ fixed and
increase $\kp_0$, moving in phase C towards phase $A$. Similarly, there 
is some evidence that $k_1(\kp_0,\Del)$ does not go to zero when 
we approach the B-C phase transition line moving in phase C, 
keeping $\kp_0$ constant while decreasing $\Del$. This points to the B-C line
being associated with UV physics. We will now discuss this further
by studying {\it lines of constant physics} in phase C.    

\section{Renormalization group flow}

The renormalization group equation in terms of 
the dimensionless gravitational coupling constant 
$\hG(a) \propto 1/k_1(a)$ reads
\beq\label{h29} 
 G(a) = a^2 \hG(a),~~~~ a \frac{\d \hG}{\d a} = -\b(\hG),~~~~
\b(\hG) = 2\hG -c \hG^2 +\dots\; .
\eeq
The $\b$-function is assumed to have a non-trivial zero
(if we ignore the dots, which signify higher-order contributions, 
it is at $\hG^* = 2/c$).
Close to the fixed point we can write
\beq\label{h30a}
\hG(a) = \hG^* - K a^\tc,~~~k_1(a) = k_1^* + \tilde{K} a^\tc,
\eeq
for some $K$, $\tilde K$, where the approach to the fixed point is governed by
the exponent
\beq\label{h30b}
\tc=-\b'(\hG^*).
\eeq
In a  standard lattice theory one would now relate
the lattice spacing near the fixed point to the bare coupling constants 
with the help of some correlation length $\xi$. 
As we have already discussed, the concept of a correlation length 
makes perfect sense even in a theory of fluctuating geometries
(see eq.\ \rf{h8} for a  definition of a diffeomorphism-invariant 
correlator). Having such a correlation length available would allow us 
to define a path of constant physics when we change the bare coupling 
constant (denoted by $g_0$), namely,
by insisting that the correlation length $\xi(g_0)$, 
expressed as the number of lattice spacings times the lattice length $a$,
represents a physical length, which is constant when we approach the 
fixed point $g_0^*$ where $\xi(g_0)$ diverges. Typically one has
\beq\label{j30}
\xi(g_0) = \frac{c}{|g_0-g_0^*|^\n},~~~{\rm i.e.}~~~
a(g_0) \propto |g_0-g_0^*|^\n,
\eeq
thus determining how we should scale the lattice spacing to zero 
as a function of the bare coupling constant $g_0$. 
However, in four-dimensional
quantum gravity without matter we do not yet have a 
suitable correlation length at our disposal which could play this role.
In search of an alternative, 
let us first consider the equation $V_4 = N_4 a^4$, which defines 
the dimensionful continuum four-volume
$V_4$ in terms of the number $N_4$ of four-simplices and the lattice
spacing. If we could 
consider $V_4$ as fixed, we could replace the $a$-dependence of \rf{h30a} by a
$N_4$-dependence, with the advantage that $N_4$ 
is a parameter we can straightforwardly control.
Re-expressing eq.\ \rf{h30a} in terms of $N_4$ yields
\beq\label{h31}
k_1(N_4) = k_1^* -K' N_4^{-\tc /4},
\eeq
for some $K'$. Since we can measure $k_1$, we could 
determine the flow to the fixed point.

In order to apply \rf{h31}, we need to ensure that 
we can consider $V_4$ as fixed when we change $N_4$.
For each value of coupling constants and each 
value $N_4$ one has a three-volume profile like the one shown 
in Fig.\ \ref{fig4}.  
Let us be more specific and rewrite \rf{h24} as
\beq\label{j31}
N_3(i) = \frac{3}{4\om} \tN_4^{3/4} \cos^3 
\left( \frac{i}{\om\tN_4^{1/4}} \right),
\eeq 
and for the fluctuations 
\beq\label{j32}
\del N_3(i) = \g \tN_4^{1/2} F\left(  \frac{i}{\om\tN_4^{1/4}} \right),
\eeq
where $\tN_4$ is the number of four-simplices of type (4,1) (and 
by construction the total number of tetrahedra contained in the spatial slices),
and where $F(t)$ is a function we have measured\cite{s4a}. 
$\tN_4$ is on average proportional to the total number of four-simplices,
the constant of proportionality depending somewhat on the choice of 
bare coupling constants, as already noted. $\g$ and $\om$ depend 
on the bare coupling constants $\kp_0$ and $\Del$. We can now write
\beq\label{j32a}
\frac{\del N_3(i)}{N_3(i)} = \frac{4\g \om}{3 \tN_4^{1/4}} \; 
\left(\frac{F(i/(\om\tN_4^{1/4}))}{\cos^3(i/(\om\tN_4^{1/4}))}\right).
\eeq   

Well away from the phase transition boundaries, our measurements
show that $\om$ and $\g$ are independent of $\tN_4$, when $\tN_4$ is 
not too small. Thus, increasing $\tN_4$ while {\it staying}
at a specific point $(\kp_0,\Del)$ in phase C does {\it not} 
correspond to keeping $V_4$ fixed,  
because during this process the size of the 
quantum fluctuations in the three-volume decreases 
relative to the expectation value of the three-volume according to
\beq\label{j50}
\frac{\del N_3}{N_3} \propto \frac{1}{\tN_4^{1/4}}.
\eeq

It is a reasonable requirement for a continuum limit
that the ratio \rf{j32a} be independent of $\tN_4$ for 
fixed $i/(\om \tN_4^{1/4})$. Thus, if we 
start at a given $\tN_4^{(0)}$, using coupling constants $\kp_0(0)$ and $\Del(0)$,
and change $\tN_4$, we have to change $\kp_0$ and $\Del$ such that
\beq\label{j51}
\g(\kp_0,\Del) \, \om(\kp_0,\Del) = 
\g(\kp_0(0),\Del(0)) \, \om(\kp_0(0),\Del(0))\;
\frac{\tN_4^{1/4}}{(\tN_4^{(0)})^{1/4}}
\eeq
holds. We can measure $ \om(\kp_0,\Del)$ and $\g(\kp_0,\Del)$ in phase C,
and have done so, but eq.\ \rf{j51}  is not sufficient to determine a curve 
$(\kp_0(\tN_4),\Del(\tN_4)$. We  need another condition. It is 
tempting to choose as a second condition that $\om(\kp_0,\Del)$ stay 
constant along a curve of constant $V_4$. This requirement implies 
according to \rf{j31} that all three-volume profiles along the 
curve can be scaled to coincide by identical scaling in ``time'' and 
``space'' directions, something which is compatible with our 
intuitive understanding of what it means to have constant $V_4$. 
The requirement that $\om(\kp_0,\Del) = \om(\kp_0(0),\Del(0))$
determines a curve starting at $(\kp_0(0),\Del(0))$. Moving 
along this curve, eq.\ \rf{j51} determines the 
relationship between $\g$ and $\tN_4$ according to
\beq\label{j52}
\g(\kp_0(\tN_4),\Del(\tN_4)) = \g(\kp_0(0),\Del(0)) \; 
\frac{\tN_4^{1/4}}{(\tN_4^{(0)})^{1/4}},
\eeq
{\it provided} it can be satisfied. 
 
In Fig.\ \rf{fig5} we show the lines of constant 
$\om$ as functions of $\kp_0$ and $\Del$, superimposed
on a contour plot of $\g(\kp_0,\Del)$. If 
we start out some distance from the B-C phase boundary, the 
plot indicates that it is impossible to 
flow to the B-C transition line along lines of constant $\om$ with 
increasing $\tN_4$. 
\begin{figure}[t]
\centerline{\includegraphics[width=4in]{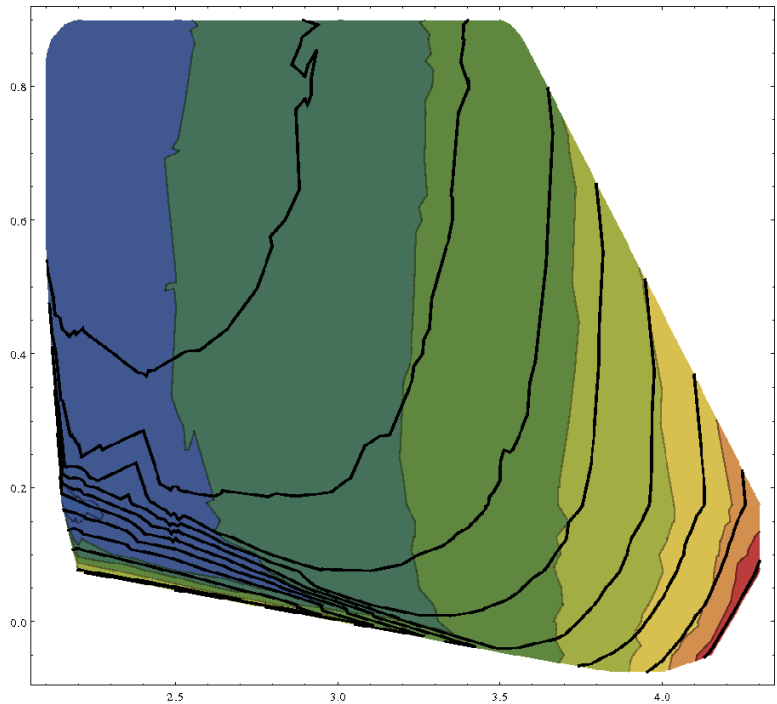}}
\caption{The flow diagram in phase C. The flow lines correspond to
constant $\om$ in eq.\ \rf{j31} (black lines). The contour plot
refers to the change of $\g$ ($\g$ increases going from the blue to the 
red region). }
\label{fig5}
\end{figure} 
However, the situation changes 
as we move closer to the B-C phase transition line. The lines
of constant $\om$ now seem to be directed towards the 
triple point. In addition, the increase of $\g$ is maximal 
when we move in the same direction towards the triple point. 
Thus there is the tantalizing possibility that we can have paths of 
constant physics leading to the triple point, with \rf{j52}
satisfied when moving along the path.  

While the condition $\om= const.$ may seem a reasonable requirement
to impose for constant $V_4$, we are not forced to adopt it. One can 
even argue that one should not use it. The reason is that $\om$
changes as a function of the bare coupling constants $\kp_0$ and $\Del$.
Its main dependence is on $\Del$, and $\om$ goes to zero 
when we approach the B-C phase transition line by decreasing 
$\Del$ (within measuring accuracy). We thus have two options:
if we regard the change in $\om$ as reflecting a change in our 
universe, the universe contracts 
in the time direction when we approach the B-C line. In order
to obtain a finite  {\it continuum} time extent at the transition
line one may therefore have to scale time and space with different 
powers of the cut-off $a$ when approaching
the B-C line. This leads naturally to a Ho\v rava-Lifshitz scenario, which we 
will discuss in the next Section. 

Alternatively, one could adopt 
the viewpoint that all universes of the form \rf{j31}
can be identified if we are allowed to rescale time relative
to space by some constant that depends on the bare coupling constants.
Such a freedom implies that we are not taking literally the microscopic 
identification of proper time as the number of lattice steps 
in the ``time'' direction times $a_t$, the lattice 
spacing in the time direction, but only claim that it is 
{\it proportional} to the ``real'' proper time.    
Thus we are not forced to take a path of constant $\om$ to 
consider $V_4$ constant. However, as mentioned above, it leaves us short of 
one condition if we want to determine a path in 
the coupling constant space corresponding to what we call
constant $V_4$.

\section{Relation to Ho\v rava-Lifshitz gravity}\label{hlgravity}

As we have described above, CDT provides a non-perturbative 
lattice regularization for a quantum theory of geometries,  
characterized by the presence of a preferred time foliation\footnote{Note that
recent work \cite{jordanloll} indicates that the strict time foliation may be relaxed, if at the
same time local causality is maintained.}. 
In the Euclidean sector one can prove that 
the lattice theory has a transfer matrix which is 
reflection-positive\cite{ajl1}. One would therefore expect that a continuum 
theory obtained from the lattice theory will have a unitary 
time evolution. 

The features just mentioned resemble those imposed in Ho\v rava-Lifshitz
gravity \cite{horava}. This suggests that CDT can provide a lattice regularization
of quantum Ho\v rava-Lifshitz gravity. Indeed, in the simplest 
case of two space-time dimensions, one can prove that CDT 
coincides with so-called quantum projectable HL gravity \cite{2dhl}
with at most quadratic derivative terms.
Of course, the situation in two dimensions is rather special. Although 1+1 dimensional 
CDT has a preferred
time foliation, it is related to standard two-dimensional Euclidean 
quantum gravity (Liouville quantum gravity) in a simple way \cite{al,ackl}.
CDT is the ``effective'' theory obtained from Liouville
quantum gravity by integrating out all Liouville quantum gravity
``baby universes". As a consequence, the amplitudes of the two theories
are related by a simple (non-analytic) mapping between their respective coupling 
constants.  

In higher dimensions, Ho\v rava's idea was to introduce 
higher-derivative terms in the spatial directions to render the 
theory renormalizable, while having only second-order time 
derivatives to ensure unitarity. Unlike  
HL gravity, CDT does not introduce explicit higher-order 
derivative terms in the spatial directions. However, as 
already mentioned, since entropic terms are important,
it may be that such higher-derivative terms are present 
in the effective CDT action. It is conceivable that for some choices of 
bare coupling constants CDT provides us with a regularization
of quantum general relativity, while for others it
is a regularization of quantum HL gravity. 

The CDT phase diagram presented in Fig.\ \ref{fig2}
has a striking similarity to a Lifshitz phase diagram, as 
noticed already elsewhere\cite{phasediagram,horavacqg}. Even the characteristics of the phases
A, B and C can be described in a language similar to the one used
for a magnetic Lifshitz system (details can be found in earlier work \cite{phasediagram}). 
 
When we move towards the second-order phase transition line B-C,
we observe a contraction of the quantum universe in the time direction.
This is in agreement with the fact that we have a second-order transition and that 
the universe in phase B collapses to a single time slice. 
It indicates that in order to obtain a genuine 
continuum limit when approaching the B-C line, one may have to scale 
time and space differently, a characteristic feature of HL gravity too.   
Various possible isotropic and anisotropic relations between 
space and time have been discussed \cite{phasediagram}. 

Can the average quantum
universe we observe on the computer  be understood as a HL universe 
rather than a universe coming from the effective GR minisuperspace
action \rf{h25}? In the context of HL gravity, one can also derive 
a minisuperspace action\cite{elias}, namely,
\beq\label{mini}
S_{mini} = \frac{2 \pi^2}{16\pi G} \int \d t \; a(t)^3 \Big( 3(1-3\lambda)\
\frac{\dot{a}^2}{a^2} -\gamma\ \frac{6}{a^2} +2 \Lambda+ \tV(a)\Big),
\eeq
where we have used the scale factor $a(t)$ instead of $V(t) \propto a(t)^3$. The
parameters $\lam$ and $\gamma$ are equal to 1 in the case of ordinary
general relativity, but can be different from 1 in HL gravity, and define the IR limit of HL gravity.
The potential $\tV(a)$ contains inverse powers of the scale factor $a$ coming from possible
higher-order spatial derivative terms.

Our reconstruction of the effective action from the computer data
is compatible with the functional form \rf{mini} of the minisuperspace action.
If we were able to extract the constant $\tilde{k}_2$ 
in front of the potential term in
\rf{h26}, it would enable us to fix the ratio $(1-3\lam)/2\g$ 
appearing in \rf{mini} \cite{s4a,semiclassical}.
At this stage, the precision of our measurements is insufficient to do so. 
The same is true for our attempts to determine $\tV(a)$ for small values 
of the scale factor, which is important for understanding 
UV quantum corrections to the potential near $a(t)=0$. 

\section{Summary}

CDT was born as an attempt to formulate a non-perturbative 
path integral directly in spacetimes with Lorentzian signature.
It can be rotated explicitly to Euclidean signature, a necessity if 
one wants to use Monte Carlo simulations to study the non-perturbatively
defined theory of fluctuating geometries. Like the ADM formalism, the Lorentzian formulation
introduces an asymmetry between space and 
time. This asymmetry survives after rotation to Euclidean signature, 
and can potentially make the theory different from so-called Euclidean
quantum gravity. CDT rotated to Euclidean signature has a positive 
definite transfer matrix and thus a unitary time evolution when 
the theory is rotated back to Lorentzian time. Both
the set-up with a preferred (proper) time and the resulting unitarity
are reminiscent of the starting point of HL gravity in the continuum. The CDT 
formalism can be used as regularization of a quantum HL gravity theory, something
which has already been done successfully in 2+1 dimensions \cite{carlip3d}. Our 
original goal was to provide a non-perturbative regularization of GR,
as a possible realization (and truly non-perturbative verification) 
of the asymptotic safety scenario. For this reason we 
have never added higher-order spatial 
derivative terms to the bare action, as is done in HL gravity, but such terms 
could in principle be created by the entropy factor. The phase diagram we 
observe is quite similar to a Lifshitz diagram, indicating that 
an interpretation in terms of HL gravity may be natural. An appealing possibility is 
that our approach includes both scenarios, where the asymptotic 
safety scenario with full GR invariance may correspond to a 
special choice of bare coupling constants, i.e.\ a fine-tuning of 
the relation between the Regge action and the entropy term in 
the effective quantum action. Better data and more observables 
will be required to discriminate between a ``pure gravity" behaviour and
an anisotropic deformation \`a la Ho\v rava-Lifshitz in the deep ultraviolet.

\subsection*{Acknowledgment}

JA and AG thank the Danish Research Council for financial
support via the grant "Quantum gravity and the role of black holes", 
and the EU for support through the ERC Advanced
Grant 291092, "Exploring the Quantum Universe" (EQU).
JJ acknowledges partial support through the International 
PhD Projects Programme of the Foundation
for Polish Science within the European Regional Development Fund of
the European Union, agreement no. MPD/2009/6. 
RL acknowledges support through several
Projectruimte grants by the Dutch Foundation for Fundamental 
Research on Matter (FOM). The contributions of JA and RL  
were supported in part by the Perimeter Institute of Theoretical Physics.
Research at Perimeter Institute is supported by the Government of Canada
through Industry Canada and by the Province of Ontario through the 
Ministry of Economic Development \& Innovation.


\begin{thebibliography}{99}

\bibitem{cambridge} 
{\it Approaches to quantum gravity}, ed. D.\ Oriti (Cambridge University Press, UK, 2009).

\bibitem{weinberg}
{S.~Weinberg},
in {\it General relativity: Einstein centenary survey}, 
eds. S.W.\ Hawking and W.\ Israel
(Cambridge University Press, UK, 1979) 790-831.

\bibitem{exactRG}
  M.~Reuter,
 {\em Phys.\ Rev.\ D} {\bf 57} 971-985 (1998);\\  
A.~Codello, R.~Percacci and C.~Rahmede,
{\it Annals Phys}.\  {\bf 324} 414 (2009);\\ 
M.~Reuter and F.~Saueressig,
arXiv:0708.1317;\\
M.~Niedermaier and M.~Reuter,
{\it Living Rev.\ Rel.}  {\bf 9} 5 (2006) 5;\\
H.W.~Hamber and R.M.~Williams,
{\it Phys.\ Rev.\  D} {\bf 72} 044026 (2005);\\
D.F.~Litim,
{\it Phys.\ Rev.\ Lett.} {\bf 92} 20130 (2004).

\bibitem{kawai}
  { H.~Kawai and M.~Ninomiya,}
  {\it Nucl.\ Phys.\ B} {\bf 336} 115 (1990);\\
  { H.~Kawai, Y.~Kitazawa and M.~Ninomiya,}
 {\it  Nucl.\ Phys.\ B} {\bf 393}   280-300 (1993);\\
 {\em  Nucl.\ Phys.\ B} {\bf 404}  684-716 (1993);
{\em Nucl.\ Phys.\ B} {\bf 467} 313-331 (1996);\\
  { T.~Aida, Y.~Kitazawa, H.~Kawai and M.~Ninomiya,}
{\em   Nucl.\ Phys.\ B} {\bf 427}  158-180 (1994). 

\bibitem{horava}
  P.~Ho\v rava,
{\it   Phys.\ Rev.\ D} {\bf 79}  084008 (2009);\\  
  P.~Ho\v rava and C.M.~Melby-Thompson,
 {\it Phys.\ Rev.\ D} {\bf 82} 064027 (2010).

\bibitem{al} J.\ Ambj\o rn and R.\ Loll,
{\it Nucl.\ Phys.\ B} {\bf 536}  407-434 (1998). 

\bibitem{ajl1}
  J.~Ambj\o rn, J.~Jurkiewicz and R.~Loll,
{\it  Nucl.\ Phys.\ B} {\bf 610} 347-382 (2001);  
{\it Phys.\ Rev.\ Lett.} {\bf 85} 924 (2000).

\bibitem{correlators}
J.~Ambj\o rn and K.N.~Anagnostopoulos,
{\it Nucl.\ Phys.\  B} {\bf 497} 445  (1997);\\ 
J.~Ambj\o rn, K.N.~Anagnostopoulos, U.~Magnea and G.~Thorleifsson,
{\it Phys.\ Lett.\  B} {\bf 388} 713 (1996);\\
J.~Ambj\o rn, J.~Jurkiewicz and Y.~Watabiki,
{\it Nucl.\ Phys.\ B} {\bf 454} 313-342 (1995);  
  J.~Ambj\o rn and Y.~Watabiki,
{\it   Nucl.\ Phys.\ B} {\bf 445} 129 (1995). 

\bibitem{regge}
T.~Regge,
{\it Nuovo Cim.}  {\bf 19}  558 (1961).

\bibitem{teitelboim}
C.~Teitelboim,
Phys.\ Rev.\ Lett.\ {\bf 50} (1983) 705-708;\\
Phys.\ Rev.\  D\ {\bf 28} (1983) 297-309.
   
\bibitem{physrep}
J. Ambj\o rn, A. G\"{o}rlich, J. Jurkiewicz, R. Loll,
Phys.\ Rep.\ {\bf 519} (2012) 127-210, 
arXiv:1203.3591.

\bibitem{ABC}
J.~Ambj\o rn, J.~Jurkiewicz and R.~Loll,
{\it Phys.\ Rev.\ Lett.} {\bf 93} 131301 (2004);  
{\it Phys.\ Rev.\ D} {\bf 72} 064014 (2005).

\bibitem{phasediagram}
J.~Ambj\o rn, A.~G\"orlich, S.~Jordan, J.~Jurkiewicz and R.~Loll,
{\it Phys.\ Lett.\ B} {\bf 690} 413-419 (2010) 413-419. 

\bibitem{jordan}
  J.~Ambj\o rn, S.~Jordan, J.~Jurkiewicz and R.~Loll,
{\it  Phys.\ Rev.\ Lett.} {\bf 107} 211303 (2011); 
{\it Phys.\ Rev.\ D} {\bf 85} 124044 (2012). 

\bibitem{s4a}
J.~Ambj\o rn, A.~G\"orlich, J.~Jurkiewicz and R.~Loll,
{\it Phys.\ Rev.\ Lett.} {\bf 100} 091304 (2008); 
{\it Phys.\ Rev.\  D} {\bf 78} 063544 (2008). 

\bibitem{semi}
J.~Ambj\o rn, J.~Jurkiewicz and R.~Loll,
{\it Phys.\ Lett.\ B} {\bf 607} 205-213 (2005). 

\bibitem{hawking}
  J.B.~Hartle and S.W.~Hawking,
{\it  Phys.\ Rev.\ D} {\bf 28} 2960-2975 (1983).

\bibitem{jordanloll} S.\ Jordan and R.\ Loll,
{\it Causal Dynamical Triangulations without preferred foliation}, arXiv:1305.4582.  

\bibitem{2dhl}
J.~Ambj\o rn, L.~Glaser, Y.~Sato and Y.~Watabiki,
{\it 2d CDT is 2d Horava-Lifshitz quantum gravity,} arXiv:1302.6359.

\bibitem{ackl}
J.~Ambj\o rn, J.~Correia, C.~Kristjansen and R.~Loll,
 {\it  Phys.\ Lett.\ B} {\bf 475} 24-32 (2000). 

\bibitem{horavacqg}
P.~Ho\v rava,
Class.\ Quant.\ Grav.\  {\bf 28} (2011) 114012, arXiv:1101.1081.
  
\bibitem{elias}
E.~Kiritsis and G.~Kofinas,
{\it Nucl.\ Phys.\  B} {\bf 821} 467  (2009) 467;\\ 
R.~Brandenberger,
{\it Phys.\ Rev.\  D} {\bf 80} 043516 (2009);\\  
G.~Calcagni,
{\it JHEP} {\bf 0909} 112 (2009).  

\bibitem{semiclassical}
  J.~Ambj\o rn, A.~Gorlich, J.~Jurkiewicz, R.~Loll, 
J.~Gizbert-Studnicki and T.~Trzesniewski,
{\it  Nucl.\ Phys.\ B} {\bf 849} 144 (2011).

\bibitem{carlip3d}
  C.~Anderson, S.~J.~Carlip, J.~H.~Cooperman, P.~Horava, R.~K.~Kommu and P.~R.~Zulkowski,
{\it  Phys.\ Rev.\ D} {\bf 85}  044027 (2012).


\end{thebibliography}

\end{document}